\documentstyle[aps,pre,multicol,psfig]{revtex}
\begin{document}
\draft
\setcounter{topnumber}{6} 
\setcounter{bottomnumber}{3}
\setcounter{totalnumber}{10} 
\renewcommand{\topfraction}{0.9}
\renewcommand{\bottomfraction}{0.8}
\renewcommand{\textfraction}{0.1}

\title{Analytic Investigations of Random Search Strategies for
  Parameter Optimization} \author{Torsten Asselmeyer and Werner
  Ebeling} \address{Institut of Physics, Humboldt University
 Berlin,\\ Invalidenstr. 110, D-10115 Berlin, Germany}
\maketitle
\begin{abstract}
  \noindent Several standard processes for searching minima of
  potential functions, such as thermodynamical strategies (simulated
  annealing) and biologically motivated selfreproduction strategies,
  are reduced to Schr\"odinger problems.  The properties of the
  landscape are encoded in the spectrum of the Hamiltonian.  We
  investigate this relation between landscape and spectrum by means of
  topological methods which lead to a possible classification of
  landscapes in the framework of the operator theory. The influence of
  the dimension $d$ of the search space is discussed.

  The connection between thermodynamical strategies and biologically
  motivated selfreproduction strategies is analyzed and interpreted in
  the above context. Mixing of both strategies is introduced as a new
  powerful tool of optimization.
\end{abstract}
\pacs{submitted to Phys.Rev. E}
\begin{multicols}{2}
\section{Introduction}
The optimization problem appears in several fields of physics and
mathematics.  It is known from mathematics that every local minimum of
a convex function defined over a convex set is also the global minimum
of the function. But the main problem is to find this optimum. From
the physical point of view every dynamical process can be considered
in terms of finding the optimum of the action functional. The best
example is the trajectory of the free point mass in mechanics which
follows the shortest way between two points.
 
Let us assume one has successfully set up a mathematical model for the
optimization problem under consideration in the form
\begin{equation}
  U(x_1,x_2,\ldots ,x_d) \longrightarrow \mbox{Min}
\end{equation}
where $U$ is a scalar potential function (fitness function) defined
over a d-dimensional vector space $X={\cal L}\{x_1,\ldots ,x_d\}$. Let
$x^{(0)}$ be the absolute Minimum of $U$ which is the search target.
Problems of this type are called parameter optimization. Typically the
search space is high-dimensional ($d\gg 1$).

The idea of evolution is the consideration of an ensemble of searchers
which move through the search space. As a illustrative example we
consider the relation between the equation of motion in mechanics,
obtained by variation of the action functional $S$ to get a trajectory
of minimal action, and the introduction of the probability
distribution for all possible trajectories by a functional integral.
Because of the weight factor $\exp(-S/\hbar)$ for every trajectory,
the trajectory of minimal action has the highest probability. The
equation of the probability distribution deduced from the functional
integral is the diffusion equation for a free classical particle. The
same idea is behind the attempt to describe optimization processes
with the help of dynamical processes.

We will be concerned with the time evolution of an ensemble of
searchers defined by a density $P(\vec{x},t)$. The search process
defines a dynamics
\begin{equation}
  P(\vec{x},t+\triangle t)=\mbox{\bf T}[P(\vec{x},t);\triangle t]
\end{equation}
with continuous or discrete time steps. An optimization dynamics
$\mbox{\bf T}$ is considered as successful, if any (or nearly any)
initial density $P(\vec{x},0)$ converges to a target density
$\lim\limits_{\tau\rightarrow\infty} P(\vec{x},\tau)$ which is
concentrated around the minimum $x^{(0)}$ of $U(\vec{x})$. We restrict
ourselves here to the case where $\mbox{\bf T}$ is given as a second
order partial differential equation. Among the most successful
strategies are the class of thermodynamic oriented
\cite{NuSa:88,An:89,SiPeHoSa:90} and the class of biological oriented
strategies \cite{Schw:95,Re:95,EbFe:89}).  Our aim is to compare on
the basis of PDE-models thermodynamical and biologically motivated
strategies by reducing both to equivalent eigenvalue problems. Further
we introduce a model for mixed strategies and investigate their
prospective power \cite{EbEn:86,BoEbEn:87,BoEb:90}.

\section{Thermodynamical Strategies}
At first we want to investigate the simplest case of an evolutionary
dynamics known in the literature as ``simulated annealing''.  The
analogy between equilibrium statistical mechanics and the Metropolis
algorithm \cite{MeRoRoTeTe:53} was first discussed by Kirkpatrick et
al. in \cite{KiGeVe:83}. There, an ensemble of searchers determined by
the distribution $P(\vec{x},t)$ move through the search space. In the
following we consider only the case of a fixed temperature. Then the
dynamics is given by the Fokker-Planck equation
\begin{equation}
  \frac{\partial}{\partial t} P(\vec{x},t)=\nabla D\cdot[\nabla
  P+\beta\nabla U P]=D\triangle P+D\nabla (\beta P\nabla U)
\label{boltzmann1}\end{equation}
with the ``diffusion'' constant $D$, the reciprocal temperature
$\beta$ and the state vector $\vec{x}$.  The stationary distribution
$P_0\sim\exp(-\beta U(\vec{x}))$ is equal to the extremum of
functional (\ref{liapunov1})\cite{Ris:89}.
 
For the case $\beta=0$ the complete analytical solution is known and
may be expressed by the well-known heat kernel or Green's function,
respectively. The corresponding dynamics is a simulated annealing at
an infinite temperature which describes the diffusion process. In this
case the optimum of the potential will be found by a random walk,
because the diffusion is not sensitive to the potential. The average
time which the process requires to move from the initial state
$\vec{x}(0)$ to the optimum $\vec{x}^{(0)}$ is given by
\begin{equation}
  t_0=\frac{1}{D}<(\vec{x}(0)-\vec{x}^{(0)})^2>
\end{equation}
We note that in several cases a generalization from the number
$D=const.$ to the case where $D$ is a symmetrical matrix is possible
\cite{Ris:89}.  Further solvable cases can be extracted from the
ansatz
\begin{equation}
  P(\vec{x},t)=\exp\left[ -\frac{\beta U(x)}{2}\right]
  y(\vec{x},t)\label{ansatz1}
\end{equation}
which after separation of the time and space variables leads to the
eigenvalue equation
\begin{eqnarray}
  (\epsilon-H)\psi(\vec{x}) &=& D\triangle\psi(\vec{x}) +
  (\epsilon-V(\vec{x}))\psi(\vec{x}) =0
\label{eigeneq1} \\
y(\vec{x},t) &=& \exp(-\epsilon t)\psi(\vec{x}) \nonumber
\end{eqnarray}
with eigenvalue $\epsilon$ and potential
\begin{equation}
  V(\vec{x}) = \frac{\beta^2}{4} D\nabla U\cdot \nabla U -
  \frac{\beta}{2} D\triangle U \label{trafo}\qquad .
\end{equation}
This equation is the well-known stationary Schr\"odinger equation from
quantum mechanics. Under the consideration of a discrete spectrum this
leads to the general solution
\begin{equation}
  P(\vec{x},t)=\exp\left[-\frac{\beta}{2}U(\vec{x})\right]
  \sum\limits_{i=0}^\infty c_i\psi_i(\vec{x})\exp(-\epsilon_i t)
  \label{generalsolution1} \qquad .
\end{equation}
To discuss more properties of equation (\ref{eigeneq1}), one has to
introduce the concept of the Liapunov functional \cite{Jet:89} defined
by the formula:
\begin{equation}
  \frac{d}{d t}y(\vec{x},t)=-\frac{\delta}{\delta y} {\cal L}(y,\nabla
  y) \label{functionalequation}\qquad .
\end{equation}
In the case of equation
\begin{equation}
  \frac{d}{d t}y(\vec{x},t)=D\triangle
  y(\vec{x},t)-V(\vec{x}))y(\vec{x},t) \qquad ,\end{equation} we
  obtain
\begin{eqnarray}
  {\cal L}(y,\nabla y) &=& \int\limits_X L(y,\nabla y) \, dvol(X)
  \label{liapunov1}\\ L(y,\nabla y) &=& \frac{D}{2}(\nabla y)^2 +
  \frac{1}{2}V y^2 \nonumber \quad .
\end{eqnarray}
For the original equation (\ref{boltzmann1}), the construction of such
functional is impossible.

We remark that the main difference between Schr\"odinger equation and
thermodynamical strategies is given by the time-dependent factor in
the solution. In quantum mechanics this set of factors $\exp(i\epsilon
t)$ forms a complete basis in the Hilbert space of functions over $t$
but in the solution of (\ref{boltzmann1}) this is not the case.

Because of the existence of an equilibrium distribution the first
eigenvalue $\epsilon_0$ vanishes and the solution is given by
\begin{equation}
  P(\vec{x},t)=\mbox{const.}\exp [-\beta U(\vec{x})] \qquad .
\end{equation}
That means that the equilibrium distribution is located around the
optimum since the exponential $\exp(-U(x))$ is a monotonous function
and the optimum is unchanged. In the limit $t\rightarrow\infty$ the
distribution $P(\vec{x},t)$ converges to the equilibrium distribution
and the strategy successfully terminates at the optimum. But this
convergence is dependent on the positiveness of the operator $H$
defined in (\ref{eigeneq1}). Usually the Laplace operator is strictly
negative definite with respect to a scalar product in the Hilbert
space of the square integrable functions ($L^2$-space). Therefore the
potential alone determines the definiteness of the operator. We thus
obtain
\[
0<\int\limits_X y(x) V(x)y(x) dvol(X) \qquad .
\]
A sufficient condition for that is
\begin{equation}
  \frac{\beta}{2}(\nabla U)^2 > \triangle U \label{condition1}
\end{equation}
which means, that the curvature of the landscape represented by
$\triangle U$ must be smaller than the square of the gradient.
Depending on the potential it thus is possible to fix a subset of $X$
on which the operator $H$ is positive definite.

Now we approximate the fitness function $U(\vec{x})$ by a Taylor
expansion around the optimum including the second order. Because the
first derivative vanishes one obtains the expression
\begin{equation}
  U(x)=U_{min}+\frac{1}{2}\sum\limits_{i=1}^d a_i
  (\vec{x}_i-\vec{x}_i^{(0)})^2 \qquad . \label{quadratic}
\end{equation}
For $a_i>0$ we get the simple harmonic oscillator which is solved by
separation of variables. The eigenfunctions are products of Hermitian
polynomials with respect to the dimension of the search space. Apart
from a constant the same result is obtained in the case $a_i<0$. A
collection of formulas can be found in appendix A.

The approximation of the general solution (\ref{generalsolution1}) for
large times leads to
\begin{equation}
  P(\vec{x},t)=c_0\exp(-\beta U(\vec{x})) + c_1e^{-\epsilon_1
    t}\exp\left[-\frac{\beta}{2}U(\vec{x})\right]\psi_1(\vec{x})+\ldots
\end{equation}
Because of the condition $\epsilon_2>\epsilon_1$, the time
$\tau=1/\epsilon_1$ can be interpreted as relaxation time, i.e. the
time for the last jump to the optimum. Even more interesting than the
consideration of the time is the calculation of velocities. One can
define two possible velocities. A first velocity $v^{(1)}$ on the
fitness landscape and a second one $v_k^{(2)}$ in the $k$-th direction
of the search space. The measure of the velocities is given by the
time-like change of the expectation values of the vector $x_k$ or the
potential $U(\vec{x})$, respectively. With respect to equation
(\ref{boltzmann1}) we obtain
\begin{equation}
  v^{(1)} = - \frac{d}{d t}<U> = D\beta <\nabla U\cdot\nabla U> -
  D<\triangle U>
\end{equation}
and
\begin{equation}
  v_k^{(2)} =- \frac{d}{d t} <x_k>= D\beta < \nabla x_k\cdot \nabla U>
  \qquad .
\end{equation} 
The velocity $v^{(1)}$ depends on the curvature and the gradient. So
we can deduce a sufficient condition for a positive velocity
\begin{equation}
  \beta(\nabla U)^2 > \triangle U \label{condition2}
\end{equation}
which is up to a factor the same as condition (\ref{condition1}). For
the quadratic potential (\ref{quadratic}) one obtains
\begin{equation}
  |x_i-x_i^{(0)}|>\frac{1}{\sqrt{a_i\beta}} \qquad \forall i \qquad .
\end{equation}
This is a restriction to a subset of $X$.

In this case it is also possible to explicitly calculate the
velocities for $a_i>0$
\begin{eqnarray}
  v^{(1)} &=& \frac{2\, c_2}{\beta \,c_1} \epsilon_2\exp(-\epsilon_2
  t) \\ v_k^{(2)} &=& \sqrt{\frac{2}{\beta a_k}} \; \frac{c_1}{c_0}\,
  \epsilon_1 \exp(-\epsilon_1 t) \qquad .
\end{eqnarray}
It is interesting to note that only the first two eigenvalues are
important for the velocities and that both velocities vanish in the
limit $t\rightarrow\infty$. Besides, the first velocity $v^{(1)}$ is
independent of the parameter $a_i$ except for special initial
conditions where the factor $c_k$ depends on the parameter $a_i$. The
other case $a_i<0$ is similar and can be found in appendix A.

\section{Fisher-Eigen Strategies}
In principle, the biologically motivated strategy is different from
the thermodynamical strategy. Whereas in the thermodynamical strategy
the population size remains constant, it is changed with respect to
the fundamental processes reproduction and selection in the case of
biologically motivated strategies, but is kept unchanged on average.
The simplest model with a similar behaviour is the Fisher-Eigen
equation given by
\begin{eqnarray}
  \frac{\partial}{\partial t}P(\vec{x},t) &=&
  [<U>-U(\vec{x})]P(\vec{x},t)+D\triangle P(\vec{x},t)
  \label{darwin}\\ <U>(t) &=& \frac{\int U(x)P(\vec{x},t) dx}{\int
    P(\vec{x},t) dx} \nonumber
\end{eqnarray}
In this case one can also form a Liapunov functional which satisfies
the equation (\ref{functionalequation}) similar to the thermodynamical
strategy. One obtains the positive functional
\begin{equation}
  {\cal L}=\int\limits_X \left( \frac{D}{2}(\nabla P)^2 -
  \frac{1}{2}(<U>-U)P^2 \right)\, dvol(X)
\end{equation}
which also has the stationary distribution as an extremum.

By using the ansatz
\begin{equation}
  P(\vec{x},t)=\exp\left[\int\limits_0^t <U>(t') dt'\right]
  y(\vec{x},t)\label{ansatz2}
\end{equation}
and the separating time and space variables, the dynamics reduces to
the stationary Schr\"odinger equation
\begin{equation}
  (\epsilon_i-H)\psi_i(\vec{x})=D\triangle\psi_i(\vec{x})+
  [\epsilon_i-U(\vec{x})]\psi_i(\vec{x})=0
  \label{eigenequation}
\end{equation}
where $\epsilon_i$ are the eigenvalues and $\psi_i(\vec{x})$ are the
eigenfunctions. This leads to the complete solution
\begin{equation}
  y(\vec{x},t)=\sum\limits_i a_i e^{-\epsilon_i t} \psi_i(\vec{x})
  \qquad .
\end{equation}
The difference to the thermodynamical strategy is given by the fact
that the eigenvalue $\epsilon_0$ in the case of the Fisher-Eigen
strategy is a non zero value, i.e. the relaxation time is modified and
one obtains
\begin{equation}
  t_0=\frac{1}{\epsilon_1-\epsilon_0} \qquad .
\end{equation}
For the harmonic potential (\ref{quadratic}) the problem is exactly
solvable for any dimension $d$ and the solution is very similar to the
thermodynamical strategy for $a_i>0$. In the other case $a_i<0$ we
obtain a different problem known from scattering theory. If the search
space is unbound the spectrum of the operator $H$ is continuous. From
the physical point of view we are interested in positive values of the
potential or fitness function (\ref{quadratic}), respectively. This
leads to a compact search space given by the interval $[-b,b]$ in each
direction with $b=\sqrt{2U_{min}/|a_i|}$. We now have to introduce
boundary conditions. The most natural choice is to let the solution
vanish on the boundary. As a result an additional restriction appears
and the spectrum of the operator $H$ is now discrete. A collection of
formulas connected to both eigenvalue problems can be found in
appendix B.

The next step is the calculation of the velocities defined in the
previous section. With respect to the Fisher-Eigen equation
(\ref{darwin}) one obtains
\begin{eqnarray}
  v^{(1)} &=& <U^2> - <U>^2 - D<\triangle U> \\ v_k^{(2)} &=& <x_k U>
  - <x_k><U> \qquad .
  \label{velocity2}
\end{eqnarray}
For the case of a quadratic potential all velocities can be calculated
from the solution and with $a_i>0$ the following expansion is
possible:
\begin{eqnarray}
  v^{(1)} &=& 4D\epsilon_0 + O(e^{(-(\epsilon_2-\epsilon_0)t)}) \\ 
  v_i^{(2)} &=& \sqrt{\frac{18D}{a_i}}\frac{c_1}{c_0}\epsilon_1
  e^{(-(\epsilon_1-\epsilon_0)t))} +
  O(e^{(-(\epsilon_3-\epsilon_0)t)}) \qquad .
  \label{velocities3}
\end{eqnarray}
These velocities are very similar to the velocities of the
thermodynamical strategy. So it is very difficult to decide whether or
not the thermodynamical strategy is faster than the biologically
motivated one or vice versa. This decision depends on the particular
circumstances. The thermodynamical strategy is faster than the
biologically motivated one in a landscape with a slight curvature and
widely extended hills whereas the biological strategy needs a
landscape with a high curvature and more localized hills to be faster
than the thermodynamical strategy.

We note the interesting fact, that there exist special tunnel effects
connected with minima of equal depth and shape. The corresponding
spectrum of the Hamiltonian shows degenerated eigenvalues. Then under
the condition of overlapping distributions located in different minima
an tunnelling with high probability between these minima is possible
(see Fig. \ref{fig:1}, dotted line). For the corresponding Boltzmann
strategy the transformed potential does not admit this tunnelling
effect (Fig. \ref{fig:1}, solid line).\\[0.3cm]
\begin{minipage}{8cm}
\begin{figure}[htbp]
  \begin{center}
    \leavevmode \psfig{figure=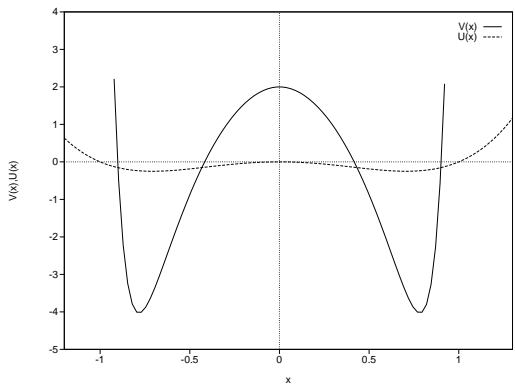,width=7cm,angle=-90}
  \end{center}
  \caption{The double well problem for Fisher-Eigen and Boltzmann}
  \label{fig:1}
\end{figure}
\end{minipage}

\section{Mixed Boltzmann--Darwin Strategies}
The dynamic equations defining Boltzmann--type search and
Fisher--Eigen type search contain a common term $D\triangle P$. Since
both types of strategies have definite advantages and disadvantages it
seems desirable to mix them. We defined the dynamics of a mixed
strategy by
\begin{eqnarray}
  \frac{\partial}{\partial t}P(\vec{x},t) &=& D\triangle
  P(\vec{x},t)+\beta D\nabla(P\nabla U)\\ &+&
  \gamma[<U>-U(\vec{x})]P(\vec{x},t)
\end{eqnarray}
For $\gamma=0$ this dynamics reduces to a pure Boltzmann strategy and
for $\beta=0$ we obtain a Fisher--Eigen strategy. The mixed case may
be treated by means of the ansatz
\begin{equation}
  P(\vec{x},t)=\exp\left[\gamma\int\limits_0^t <U>dt' - \frac{1}{2}
  \beta U(\vec{x}\right] y(\vec{x},t)
\end{equation}
which leads to the explicit solution in the eigenfunctions of the
problem
\begin{eqnarray}
  0 &=& D\triangle \psi_i(\vec{x}) + [\epsilon_i -
  E(\vec{x})]\psi_i(\vec{x}) \\ E(\vec{x}) &=& \gamma
  U-\frac{D\beta}{2}\triangle U +\frac{D\beta^2}{4}(\nabla
  U)\cdot(\nabla U) \nonumber \qquad .
\end{eqnarray}
For $\beta=0$ and $\gamma=1$ we end up with the case of Fisher-Eigen
strategies and for $\gamma=0$ , $\beta>0$ the Boltzmann dynamics is
obtained. In this respect the mixed strategy is indeed a
generalization of both cases (see Fig. \ref{fig:2}).\\[0.3cm]
\begin{minipage}[]{8.5cm}
\begin{figure}[htbp]
  \begin{center}
    \leavevmode
    \psfig{figure=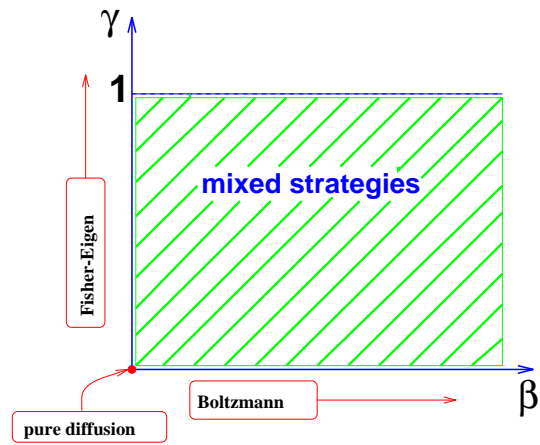,width=7cm}
  \end{center}
  \caption{The parameter dependence of the mixed strategy}
  \label{fig:2}
\end{figure}
\end{minipage}\\[0.3cm]
The linearity of the differential equation leads to simple relations
between the solutions and velocities. For instance the velocities for
the thermodynamical and the biological strategy can be added with
respect to the constant $\gamma$ to get the velocities of the mixed
strategy. For the solution of the problem one simply takes the
solution of the thermodynamical strategy (\ref{solution1}) and
redefines the coordinate $\xi_k$ by
\[ \xi_k = x_k \sqrt[4]{\frac{\beta^2\, a_k^2}{4}+\frac{\gamma\, a_k}{2D}} 
\qquad .\] So the mixed strategy combines the advantages of both
strategies if one uses a criterion to choose the coefficients $\gamma$
and $\beta$.  The dependence on the 3 optimization parameters $D$,
$\beta$ and $\gamma$ is a rather simple one. For a given $D$, the best
strategy is evidently an increase of the $\beta$-value, since the
$\gamma$-dependence is weak. However, if the ``friction'' $\beta D$ is
fixed by the conditions, the maximization of $\gamma/\beta$ leads to
the smallest relaxation time. In other words, if the loss $\beta D$ is
fixed, maximal $\gamma$ (i.e. most competition) and minimal $\beta$
(i.e. large temperature) is the best choice for a fast search of the
minimum. In general the optimal search requires $\beta>0$ and
$\gamma>0$. With other words, adding some amount of the
``complementary'' strategy is in most cases to be recommended. This
was already found empirically in a earlier work \cite{EbEn:86,BoEbEn:87}. 

\section{Global properties of Strategies}
The study of the problem given by the potential (\ref{quadratic}) is
the same as the study of the local properties of the landscape. It is
known from the Lemma of Morse, that there is a coordinate system so
that every function with non-degenerate critical points in the
neighborhood $U_p$ of one critical point $p$ can be expressed as
\[ f= f(p)-(y_1)^2-\ldots -(y_{\lambda})^2 + (y_{\lambda+1})^2+\ldots 
+(y_d)^2 \]
with $(y_1,\ldots ,y_d)$ as coordinate system. The number $\lambda$ is
the index of the critical point. Now we use the compact space
$X\cup\{\infty\}$ rather than $X={\mbox{\rm I\kern -.2em R}}^d$ which
in contrast to ${\mbox{\rm I\kern -.2em R}}^d$ is homeomorphic to
$S^d$, the $d$-dimensional sphere. Now every function over $S^d$ is
non-degenerate and we obtain for the number $C_\lambda$ of critical
points with index $\lambda$ the Morse inequalities \cite{Mil:63}
\begin{eqnarray}
  1\le C_0 \qquad 1 &\le& C_d \qquad 0\le C_\iota \\
1+(-1)^d &=& \sum\limits_{\lambda} (-1)^\lambda C_\lambda \label{morse1}
\end{eqnarray}
with $\iota=1,\ldots d-1$. So we find that in the case of many maxima
and minima the number of saddle points increases \cite{conebe:92}.
This is a global statement which only depends on the topology of the
sphere $S^d$.  

In the case of a compact, simple-connected subset of ${\mbox{\rm
    I\kern -.2em R}}^d$ one obtains a similar result because the
boundary of this subset is homeomorphic to a $d-1$ dimensional sphere
and the Morse inequalities can be used again. If this subset has no
boundary then we obtain
\begin{equation}
1=\sum\limits_{\lambda} (-1)^\lambda C_\lambda \label{morse2}
\end{equation}
and the result of \cite{conebe:92} can be deduced again. The treatment
of constraints in the optimization problem is connected with the
occurrence of holes in the space. This leads to a correction of the
formulas (\ref{morse1}) or (\ref{morse2}), respectively. The number of
saddle points increases with respect to the number of holes.

Next we want to study the influence of the landscape on the solutions
in terms of the dynamics, which in our case is restricted to the
influence of the potential or fitness function, respectively. The
general form of this dynamics is given by
\begin{equation}
\frac{\partial}{\partial t}P(x,t)=-HP(x,t)
\end{equation}
with the selfadjoint Operator
\begin{eqnarray}
  H_B &=& -D\nabla\cdot (\nabla + \beta\nabla U)
  \qquad\mbox{Boltzmann} \\ H_{FE} &=& -<U>+U-D\triangle
  \qquad\mbox{Fisher-Eigen}\qquad .
\end{eqnarray}
This operator equation has the formal solution
\begin{equation}
P(x,t)=\exp(-tH)P(x,0)
\end{equation}
acting on the initial condition. With the help of (\ref{ansatz2}), the
operator for the Fisher-Eigen strategy can be reduced to
$H=U-D\triangle$. Then both operators are in the same class known as
generalized Laplacian. The representation of this class from a unified
point of view is possible. To this end we introduce the Dirac operator
as a first order differential operator and regard the generalized
Laplacian $H$ as the square of the Dirac operator. It was shown in
\cite{bvg:90} that every generalized Laplacian can be represented by
the square of a Dirac operator with respect to a suitable Clifford
multiplication. For the two cases one obtains
\begin{eqnarray}
  D_B &=& i\gamma^\mu(\partial_\mu + \partial_\mu U)
  \qquad\mbox{Boltzmann} \\ D_{FE} &=& \gamma^\mu(\partial_\mu +
  iA_\mu(x))\qquad\mbox{Fisher-Eigen}
\end{eqnarray}
where $\gamma^\mu$ are the Dirac matrices ($\mu=1,\ldots d$) and
$A_\mu(x)$ is a locally defined vector field with 
$[\gamma^\mu,\gamma^\nu]\partial_\mu A_\nu (x)=-i U(x)$. Together with
the expression
\begin{equation}
  D=\left(\begin{array}{cc} 0 & {/\mkern -10mu \partial} \\ {/\mkern
    -10mu \partial}^* & 0 \end{array}\right)
\end{equation}
for every Dirac operator $D$, one can simple calculate the squares of
the Dirac operators to establish
\begin{equation}
  H_B= D_B^2=\left(\begin{array}{cc} {/\mkern -10mu
    \partial}_B{/\mkern -10mu \partial}_B^* & 0 \\ 0 & {/\mkern -10mu
    \partial}_B^*{/\mkern -10mu \partial}_B \end{array}\right) \qquad
  H_{FE} = D_{FE}^2
\end{equation}
with the adjoint operator ${/\mkern -10mu \partial}^*$. This means
that both problems can be described by the motion of a fermion in a
field $\nabla U$ or $iA(x)$, respectively. The equilibrium state (or
stationary state) is given by the kernel of the operator $H$ which is
the direct sum $\ker H=\ker{/\mkern -10mu \partial}\oplus\ker{/\mkern
  -10mu \partial}^*$. If the dimension $d$ is even, which is always
true for the high-dimensional case, this splitting can be introduced
by the product of all $\gamma$-matrices usually denoted by $\gamma^5$.
Because of the compactness of the underlying space $S^d$, the kernel
of $H$ is finite-dimensional and the spectrum is discrete. We note,
that the spectrum of both, ${/\mkern -10mu \partial}{/\mkern -10mu
  \partial}^*$ and ${/\mkern -10mu \partial}^*{/\mkern -10mu
  \partial}$, is equal up to the kernel. So the interesting
information about the problem is located in the asymmetric splitting
of the kernel $\ker H$. The physical interpretation is given by an
asymmetric ground state of the problem which is only connected to the
geometry of the landscape. In more mathematical terms both Dirac
operators are described as covariant derivatives with a suitable
connection. Together with the fiber bundle theory \cite{Hus:66} and
the classification given by the K-theory \cite{Ati:67} one obtains a
possible classification of the fitness landscapes in dependence of
convergence velocities given by the periodicity of the real K-theory with
period 8.  Each of these 8 classes describes a splitting of the kernel
$\ker H$ and leads to a different velocity. A complete description of
this problem will be published later on.

\section{Conclusions}
In physics the most classical dynamical processes follow the principle
of minimization of a physical quantity which often leads to an
extremum of the action functional. This problem frequently has a
finite number of solutions given by the solutions of a differential
equation also known as the equation of motion.  Investigating this
fact in relation to optimization processes, one obtains in the
simplest case the thermodynamical and the biological strategy. The
description is given by the distribution of the searcher and a
dynamics of the distribution converging to an equilibrium distribution
located around the optimum of the optimization problem. With the help
of the kinetics and the eigenfunction expansion, we investigated both
strategies in view of the convergence velocity. In principle both
strategies are equal because one obtains a stationary Schr\"odinger
equation. But, the main difference is the transformation of the
fitness function (or potential) from $U(x)$ to $V(x)$ (see
(\ref{trafo})) in the case of the thermodynamical strategy. The
difference of both strategies leads to the idea of adding a small
amount of the ``complementary'' strategy in order to hope for an
improvement. The difference in the velocity on the one hand and the
similarity in the equation on the other look like a unified treatment
of both strategies under consideration. This is represented in the
last section in the formalism of fiber bundles and heat kernels to get
the interesting result, that up to local coordinate transformations
the strategies are split into 8 different classes.
\end{multicols}

\appendix
\section*{A - Thermodynamical strategy}

For the case of quadratic potentials (\ref{quadratic}) the problem
(\ref{boltzmann1}) may be solved explicitly. We get the eigenvalues
\[ \epsilon_{n_1\ldots n_d}=U_{min}+\sum\limits_{i=1}^d a_i \beta D
\left( n_i+\frac{1}{2}\right) \qquad n_i=0,1,2,\ldots \]
and the eigenfunctions
\begin{eqnarray*}
  \psi_{n_1\ldots n_d}(x_1,\ldots ,x_d)=\prod\limits_{i=1}^d
  \psi_{n_i}(x_i) \\ \psi_{n_i}(x_i)=\exp \left( -\frac{\beta a_i}{4}
  x_i^2 \right) H_{n_i}\left( \sqrt{\frac{\beta a_i}{2}} x_i \right)
\end{eqnarray*}
which lead to the solution
\begin{equation}
  P(\vec{x},t)=\sum\limits_{i=n_1+\ldots +n_d} c_i
  \prod\limits_{k=1}^d \left[\exp (-\xi_k^2) H_{n_k}(\xi_k) \right]
  \exp(-\epsilon_i t) \label{solution1}
\end{equation}
with $\xi_k=x_k\, \sqrt{\beta a_k/2}$. Next we have to fix initial
conditions for this problem. At first one starts with a strong
localized function, i.e. with a delta distribution.
\[ P(\vec{x},0)=\prod\limits_{i=1}^d \delta(x_i - x^{(0)}_i) \]
Because of the relation:
\begin{equation}
  \int \exp(-\xi^2)H_n(\xi)H_m(\xi) d\xi = \delta_{mn} \int
  \exp(-\xi^2)H^2_n(\xi) d\xi =\delta_{mn} N_n = \delta_{mn} 2^n (n!)
  \sqrt{\pi}
\end{equation}
we obtain for the coefficients
\begin{equation}
  c_k = \prod\limits_{i=1}^d \frac{2}{N_{n_i}\beta a_i} H_{n_i}\left(
  \sqrt{\beta a_i}{2}x^{(0)}_i\right) \qquad .
\end{equation}
with $k=n_1 + n_2 + \ldots + n_d$.  In the case of the full symmetry
$a=a_1=a_2=\ldots =a_d$ we can calculate the radial problem to obtain
the eigenvalues
\[ \epsilon = a\beta Dd \left( k+\frac{1}{2}\right) \]
with $d$ as the dimension of the landscape.  The calculation of the
velocities leads to two cases for the potential (\ref{quadratic}):
\begin{enumerate}
\item $a_i>0$ \begin{eqnarray} v^{(1)} &=& \frac{2\, c_2}{\beta \,c_1}
  \epsilon_2\exp(-\epsilon_2 t) \\ v_k^{(2)} &=& \sqrt{\frac{2}{\beta
      a_k}} \; \frac{c_1}{c_0}\, \epsilon_1 \exp(-\epsilon_1 t)
\end{eqnarray}
\item $a_i<0$ \begin{eqnarray} v^{(1)} &=& \frac{2}{\beta N_1}
  \sum\limits_{k=0}^\infty c_{2k}\exp(-\epsilon_{2k}t)\epsilon_{2k}
  \sum\limits_{i=1}^d
  \frac{H_{2k+3}(b_i)}{(2k+1)(2k+2)(2k+3)}\prod\limits_{j=1\, i\not= j
    \atop k_1+\ldots k_d=k}^d\frac{H_{2k_i+1}(b_j)}{(2k+1)}\\ 
  v_i^{(2)} &=& \sqrt{\frac{2}{\beta a_k}} \; \frac{1}{N_1}
  \sum\limits_{k=0}^\infty
  c_{2k+1}\exp(-\epsilon_{2k+1}t)\epsilon_{2k+1}
 \frac{H_{2k+3}(b_i)}{(2k+2)(2k+3)}\prod\limits_{j=1\,
    i\not= j \atop k_1+\ldots
    k_d=k}^d\frac{H_{2k_i+1}(b_j)}{(2k+1)}\end{eqnarray}
\end{enumerate}
where 
\[ N_1= \sum\limits_{k=0}^\infty c_{2k}\exp(-\epsilon_{2k}t)
\prod\limits_{i=1 \atop k_1+\ldots k_d=k}^d\frac{H_{2k_i+1}(b_i)}{(2k+1)}\]
is the normalization factor and $b_i=\sqrt{2U_m/a_i}$ is the interval
length.

\section*{B - Biological strategy}

For a harmonic potential (\ref{quadratic}) the problem (\ref{darwin})
is exactly solvable for any dimension $d$. We get the eigenvalues
\[ \epsilon_{n_1\ldots n_d}=U_{min}+\sum\limits_{i=1}^d \sqrt{2a_i\, D}
\left( n_i+\frac{1}{2}\right) \qquad n_i=0,1,2,\ldots \]
and the eigenfunctions
\begin{eqnarray*}
  \psi_{n_1\ldots n_d}(x_1,\ldots ,x_d)=\prod\limits_{i=1}^d
  \psi_{n_i}(x_i) \\ \psi_{n_i}(x_i)=\exp \left(
  -\sqrt{\frac{a_i}{4D}} x_i^2 \right) H_{n_i}\left(
  \sqrt[4]{\frac{a_i}{2D}} x_i \right)
\end{eqnarray*}
which lead to the solution
\begin{equation}
  P(\vec{x},t)=\sum\limits_{i=n_1+\ldots +n_d} c_i
  \prod\limits_{k=1}^d \left[\exp \left(-\frac{\xi_k^2}{2}\right)
  H_{n_k}(\xi_k) \right] \exp(-\epsilon_i t)
\end{equation}
with $\xi_k=x_k\, \sqrt[4]{a_k/(2D)}$.

We now come to the problem of the maximum, i.e. the potential
(\ref{quadratic}) with $a_i<0$.  The solution for one dimension
(direction $i$) is simply obtained as
\begin{equation}
\psi_{\lambda_i}(x_i)=D_{-iB_i-\frac{1}{2}}\left(\sqrt[4]{\frac{a_i}{2D}}\;
e^{-i\pi/4} x_i \right)
\end{equation}
with $B_i=\lambda_i/(2\sqrt{D a_i})$ and $D_k(x)$ as parabolic Bessel
functions depending on the parameter $B_i$. In practice one is
interested in positive values of the fitness function or potential
which leads to a restriction of the search space $X$ to the a
hypercube with length $b_i=\sqrt{2U_m/a_i}$ in every dimension. We
claim that the solution vanishes at the boundary of the hypercube.
This restriction leads to a discrete spectrum. The zeros of the
parabolic Bessel function are given by the solution of the equation
\begin{equation}
\frac{1}{4}\left(\rho_{ki}\sqrt{\rho_{ki}^2-1}-\mbox{arccosh}\,
\rho_{ki}\right)=\frac{(-A_n)^{3/2}}{6\sqrt{\frac{U_m+n_{i}}{2\sqrt{a_iD}}}}
\end{equation}
where the coefficients $A_i$ can be found in the book \cite{Math:72}
and the zeros are defined by $c=2\sqrt{a_i}\rho$. For the eigenvalues
$\epsilon_k$ in (\ref{eigenequation}) one obtains
\begin{equation}
  \epsilon_n = \sum\limits_{i=1 \atop n_1+\ldots n_d=n}^{d} n_i
  \qquad\qquad n_i = \frac{U_m}{\rho_{ki}^2}\sqrt{\frac{D}{a_i}} - U_m
  \qquad .
\end{equation}

The calculation of the velocities is very lengthy and one obtains
\begin{enumerate}
\item $a_i>0$ (minimum):
\begin{eqnarray}
  v^{(1)} &=& \frac{1}{N_1} \sum\limits_{i=1}^d \sqrt{\frac{D
      a_i}{2}}\left( c_0\epsilon_0e^{-\epsilon_0
  t}+\sum\limits_{k=1}^\infty
  (-1)^{k+1}\frac{(2k)!}{k!}(2k^2+k+\frac{3}{2})c_{2k}\epsilon_{2k}e^{-\epsilon_{2k}
  t}\right) \\ v_i^{(2)} &=& \sqrt{\frac{2D}{a_i}}\frac{3}{N_1} \left(
  \sum\limits_{k=0}^\infty (-1)^{k}\frac{(2k+1)!}{k!}
  c_{2k+1}\epsilon_{2k+1}\exp(-\epsilon_{2k+1} t)\right)
\end{eqnarray}
with
\begin{equation}
  N_1=\sum\limits_{k=0}^\infty
  (-1)^{k}\frac{(2k)!}{k!}c_{2k}\exp(-\epsilon_{2k} t)
\end{equation}
as normalization.
\item case $a_i<0$ (maximum):
\begin{eqnarray}
  v^{(1)} &=& \frac{1}{N_2} \sum\limits_{n=1 \atop n_1+\ldots
    n_d=n}^{\infty}\epsilon_n e^{\epsilon_n t}\prod\limits_{i=1}^d
  \left(\sqrt{\left|
      \frac{\Gamma(\frac{1}{4}+ib_i/2)}{\Gamma(\frac{3}{4}+ib_i/2)}
  \right|}\right)
    \sum\limits_{k=1}^d I_k(\xi_k^2)\sqrt{\frac{D\, |a_k|}{2}} \\ 
    v_k^{(2)} &=&\frac{1}{N_2} \sum\limits_{n=1 \atop n_1+\ldots
      n_d=n}^{\infty}\epsilon_n e^{\epsilon_n t}\prod\limits_{i=1}^d
    \left(\sqrt{\left|
        \frac{2\Gamma(\frac{3}{4}+ib_i/2)}{\Gamma(\frac{1}{4}+ib_i/2)}\right|
        }I_i()\right)I_k(\xi_k)\sqrt[4]{\frac{|a_k|}{2D}}
\end{eqnarray}
with 
\begin{eqnarray}
  I_k(\xi_k) &=& \int\limits_{-u_k}^{u_k} \xi_k y_1(\xi_k) d\xi_k
  \qquad I_k(\xi_k^2)=\int\limits_{-u_k}^{u_k} \xi_k^2 y_2(\xi_k)
  d\xi_k \qquad I_k()=\int\limits_{-u_k}^{u_k} y_1(\xi_k) d\xi_k \\ 
  u_k &=& \frac{2U_m}{|a_k|}\sqrt[4]{\frac{|a_k|}{2D}} \qquad \qquad
  b_k=\frac{n_k}{2\sqrt{D |a_k|}} \\ N &=& \sum\limits_{n=1 \atop
    n_1+\ldots n_d=n}^{\infty} e^{\epsilon_n t}\prod\limits_{i=1}^d
  \left(\sqrt{\left|
      \frac{\Gamma(\frac{1}{4}+ib_k/2)}{\Gamma(\frac{3}{4}+ib_k/2)}\right|}\,
    I_k()\, \right)
\end{eqnarray}
and functions $y_1(), y_2()$ defined in \cite{Math:72} page 692 as
series
\begin{eqnarray}
  y_1(\xi_k) &=& 1 +
  b_k\frac{\xi_k^2}{2!}+\left(b_k^2-\frac{1}{2}\right)\frac{\xi_k^4}{4!}+\ldots
  \\ y_2(\xi_k) &=& \xi_k +
  b_k\frac{\xi_k^3}{3!}+\left(b_k^2-\frac{3}{2}\right)\frac{\xi_k^5}{5!}+\ldots
\end{eqnarray} 
\end{enumerate}
\begin{multicols}{2}

\end{multicols}


\begin{references}

\bibitem{NuSa:88}
J.D. Nulton and P.~Salamon.
\newblock Statistical mechanics of combinatorical optimization.
\newblock {\em Phys. Rev.}, {\bf A 37}:1351, 1988.

\bibitem{An:89}
B.~Andresen.
\newblock Finite-time thermodynamics and simulated annealing.
\newblock In {\em Proceedings of the Fourth International Conference on
  Irreversible Processes and Selforganization}, Rostock, 1989.

\bibitem{SiPeHoSa:90}
P.~Sibiani, K.M. Pedersen, K.H. Hoffmann, and P.~Salamon.
\newblock Monte Carlo dynamics of optimization: A scaling description.
\newblock {\em Phys. Rev.}, {\bf A 42}:7080, 1990.

\bibitem{Schw:95}
H.P. Schwefel.
\newblock {\em Evolution and optimum seeking}.
\newblock Wiley, New York, 1995.

\bibitem{Re:95}
I.~Rechenberg.
\newblock {\em Evolutionsstrategien - Optimierung technischer Systeme nach
  Prinzipien der biologischen Information}.
\newblock Friedrich Frommann Verlag (G\"unther Holzboog K.G.), Stuttgart - Bad
  Cannstatt, 1995.

\bibitem{EbFe:89}
R.~Feistel and W.~Ebeling.
\newblock {\em Evolution of Complex Systems}.
\newblock Kluwer Academic Publ., Dordrecht, 1989.

\bibitem{EbEn:86}
W.~Ebeling and A.~Engel.
\newblock Models of {E}volutionary systems and their applications to
  optimization problems.
\newblock {\em Syst. Anal. Model. Simul.}, {\bf 3}:377, 1986.

\bibitem{BoEbEn:87}
T.~Boseniuk, W.~Ebeling, and A.~Engel.
\newblock Boltzmann and {D}arwin strategies in complex optimization.
\newblock {\em Phys. Lett.}, {\bf A 125}:307, 1987.

\bibitem{BoEb:90}
T.~Boseniuk and W.~Ebeling.
\newblock Boltzmann- , {D}arwin- and {H}eackel-strategies in optimization
  problems.
\newblock In {\em PPSN Dortmund}, 1990.

\bibitem{MeRoRoTeTe:53}
N.~Metropolis, A.~Rosenbluth, M.~Rosenbluth, A.~Teller, and E.~Teller.
\newblock {\em J. Chem. Phys.}, {\bf 21}:1087, 1953.

\bibitem{KiGeVe:83}
S.~Kirkpatrick, C.D. Gelatt~Jr., and M.P. Vecchi.
\newblock {\em Science}, {\bf 220}:671, 1983.

\bibitem{Ris:89}
H.~Risken.
\newblock {\em The {F}okker-{P}lanck Equation}.
\newblock Springer Verlag, 1989.

\bibitem{Jet:89}
G.~Jetschke.
\newblock {\em Mathematik der {S}elbstorganisation}.
\newblock Deutscher Verlag der Wissenschaften, 1989.

\bibitem{Mil:63}
J.~Milnor.
\newblock {\em Morse theory}.
\newblock Ann. of Mathematical Study {\bf 51}. Princeton Univ. Press,
  Princeton, 1963.

\bibitem{conebe:92}
M.~Conrad and W~Ebeling.
\newblock M.{V}. {V}olkenstein, evolutionary thinking and the structue of
  fitness landscapes.
\newblock {\em BioSystems}, {\bf 27}:125, 1992.

\bibitem{bvg:90}
N.~Berline, M.~Vergne, and E.~Getzler.
\newblock {\em Heat kernels and Dirac Operators}.
\newblock Springer Verlag, New York, 1992.

\bibitem{Hus:66}
D.~Husemoller.
\newblock {\em Fiber Bundles}.
\newblock MacGraw-Hill Book Co., New York-London-Sydney, 1966.

\bibitem{Ati:67}
M.~Atiyah.
\newblock {\em K-Theory}.
\newblock W.A. Benjamin, New York--Amsterdam, 1967.

\bibitem{Math:72}
Milton Abramowitz and Irene~A. Stegun.
\newblock {\em Handbook of Mathematical Functions with Formulas, Graphs, and
  Mathematical Tables}.
\newblock Applied Mathematical Series 55. National Bureau of Standards,
  December 1972.

\end{references}
\end{document}